\documentclass[10pt,journal]{IEEEtran}
\usepackage{amsmath,amsfonts}
\usepackage{array}
\usepackage{textcomp}
\usepackage{stfloats}
\usepackage{url}
\usepackage{verbatim}
\usepackage{graphicx}
\usepackage{cite}
\usepackage{amssymb,amsfonts}
\usepackage{cases}
\usepackage{graphicx}
\usepackage{xcolor}
\usepackage{bm}

\def\BibTeX{{\rm B\kern-.05em{\sc i\kern-.025em b}\kern-.08em
    T\kern-.1667em\lower.7ex\hbox{E}\kern-.125emX}}

\usepackage{algorithm}
\usepackage{amsmath}
\usepackage{algpseudocode}

\usepackage{booktabs}
\usepackage{multirow}  

\usepackage{amsfonts}
\usepackage{color} 
\usepackage{subcaption}
\usepackage{amsmath}
\usepackage{amsthm}

\usepackage{balance}

\begin{document}
\title{  Beyond Profit: A Multi-Objective Framework for Electric Vehicle   Charging Station Operations }

\author{\IEEEauthorblockN{ Shuoyao Wang}  and \IEEEauthorblockN{Jiawei Lin} \\
College of Electronic and Information Engineering, Shenzhen University, China \\
E-mail: sywang@szu.edu.cn, linjiawei2021@email.szu.edu.cn 
\thanks{Shuoyao Wang is the corresponding author. This work is supported in part by the National Natural Science Foundation of China (Project number 62101336); in part by the Guangdong Basic and Applied Basic Research Foundation (Project number 2022A1515011301).}
}

% \markboth{Journal of \LaTeX\ Class Files,~Vol.~18, No.~9, September~2020}%
% {How to Use the IEEEtran \LaTeX \ Templates}

\maketitle
\providecommand{\keywords}[1]{\textbf{\textit{Index terms---}} #1}
\pagestyle{empty}  % no page number for the second and the later pages
\thispagestyle{empty} % no page number for the first page
\begin{abstract}
 This paper explores the pricing and scheduling strategies of the  electric vehicle charging stations in response to the rising demand for cleaner transportation. Most of the existing methods focus on maximizing the
energy efficiency  or the charging station profit, however, the reputation of EVs is also
a key factor for the long-term charging station operations.   To address these gaps, we propose a novel framework for jointly optimizing pricing and continuous-multiple charging rates. Our approach aims to maximize both charging station profit and reputation, considering multi-objective optimization and continuous rate control within physical constraints. Introducing a pricing fluctuating penalty for reputation modeling and a linear programming-based safe layer for constraints, we confront the complexity of continuous charging rates' action space. To enhance convergence, we explore a soft action critic framework with novel entropy temperature tunning technique. %In summary, this paper contributes by formulating and tackling the challenges of pricing and continuous-multiple charging rates scheduling, with a focus on maximizing profit and reputation for EV charging stations. 
The experiments conducted with real data demonstrate that  the  proposed method can provide extra  25.45\%-52.20\% average JPR than the representative baselines. \end{abstract}
\section{Introduction}
The utilization of electric vehicles (EVs) is on the rise, driven by advancements in battery technology and government support, with a focus on mitigating air pollution and addressing health concerns. Globally, countries are implementing fuel regulations that underscore the distribution of EVs as a cleaner mode of transportation. The proliferation of EV charging stations has played a pivotal role in aligning energy demand and supply~\cite{7786114}. Serving as intermediaries, these stations influence energy flexibility and efficiency through pricing and scheduling strategies. %To integrate EVs into distribution networks efficiently, research focuses on charging station pricing and scheduling.
%The use of electric vehicles (EVs) is growing due to battery
%technology advancements and supportive government policies.
%EVs are considered a solution to air pollution and fine dustrelated health issues. Countries worldwide are implementing
%fuel-usage regulations and focusing on EV distribution and
%expansion to establish cleaner transportation systems.  
%With the development of EVs, the number of EV charging stations has also significantly increased \cite{7786114}. AS intermediaries between EVs and electricity utility companies, charging stations play an important role in coordinating energy demand and supply, such as demand response pricing and charging scheduling \cite{7056190,6887361,shao2023preemptive,9524742}. Indeed, the joint pricing and scheduling strategy of charging stations is crucial for improving the charging flexibility and charging efficiency \cite{9764664}.  
%However, High penetration of EVs is expected
%to change the power load profile significantly in distribution networks, creating potential threats to the power grid \cite{5356176,9495175,7786114}. Establishing a conveniently available public charging infrastructure is
%essential to accommodating more clean energy, reducing carbon
%emissions, and alleviating peak charging loads. In the past
%decade, various EV charging control and scheduling schemes
%have been proposed to improve grid reliability \cite{7056190}, reduce
%charging operation cost \cite{6887361}, offer auxiliary services \cite{shao2023preemptive}, and
%promote the integration of renewable generation in commercial
%Microgrids \cite{9524742}, etc. 

To accommodate a higher penetration of EVs safely and efficiently in today's distribution networks, extensive research efforts have been devoted to the pricing and scheduling problem of charging station. 
Existing work can be broadly categorized into two groups: model-based optimization and reinforcement learning (RL) methods. 
%The existing work can be divided into two categories: model-based optimization and reinforcement learning (RL) methods. For instance, %\cite{li2022accessible} proposed a game-based charging scheduling strategy incorporating energy and information interaction between EVs to facilitate charging scheduling. 
%Integrated the charging scheduling with pricing scheme, \cite{kapoor2022centralized} proposed three centralized and decentralized pricing schemes for day-ahead EV scheduling, aiming to maximize the profit of a distribution system operator. To minimize EV charging costs, \cite{mohammed2022interruptible} put forth an interruptible charging schedule scheme and assessed its effectiveness in residential and commercial CSs with varying time-of-use pricing structures.
For example,  \cite{li2022accessible} proposed a game-based charging strategy for EV interaction. Ref.~\cite{kapoor2022centralized} integrated pricing with scheduling, proposing centralized and decentralized schemes.  Ref.~\cite{mohammed2022interruptible} suggested an interruptible charging schedule to minimize costs and assess effectiveness. However, many model-based methods assume prior knowledge, which may not always hold true in real-world scenarios, leading to suboptimal outcomes.
%However, most model-based methods assume the charging stations have the prior knowledge about EV and utility behavior patterns. Such assumption can not always holds in the real world.  Moreover, the relaxation technique to solve the optimization problem leads to the suboptimality of the model-based methods. 

To address the above issues,  RL methods have been increasingly employed to develop scheduling and pricing schemes in recent years. For instance,  \cite{8888199}  explored reinforcement learning with feature engineering  to learn the joint pricing and charging scheduling strategy. Subsequent research has delved into utilizing advanced machine learning algorithms for improved decision-making, such as actor critic framework~\cite{9444352}, long and short-term memory~\cite{9766186}, and multi-agent RL~\cite{9764664}. More recently, efforts have extended RL methods to scenarios involving continuous charging rates~\cite{9653670} and vehicle-grid integration~\cite{9750198} scenarios. 
 Considering the charging network with multiple charging stations, \cite{liu2022collaborative} proposed a two-layer optimization method for charging scheduling, and \cite{shin2019cooperative}   employed multi-agent deep RL (MA-DRL) to determine the energy to be charged or discharged by cooperative charging stations.

Nevertheless, there are two main shortcomings limiting the application of the exciting Rl-based studies in the real electricity market.
\begin{itemize}
	\item Most methods focus solely on maximizing energy efficiency from the utility perspective or charging station profit from the station's viewpoint. User satisfaction is often treated as binary, fulfilling the charging demand before the deadline or not. In addition to energy efficiency and station profit, the reputation of EVs is a crucial factor for long-term charging station operations \cite{9124670,9095998}.
	\item To ensure RL convergence, existing methods often make relaxations, such as i) discretizing charging rates into binary/discrete decisions ~\cite{9766186}\cite{9750198}\cite{10323291}, ii) eliminating the total charging rate capacity constraints~\cite{9766186}\cite{9653670}\cite{9750198}, or iii) simply determining the total charging rate and then using heuristic dispatch to allocate total energy to each EV~ \cite{8888199}\cite{9444352}\cite{9764664}\cite{10323291}. While these relaxations solve the convergence problem for complex action spaces, they often lead to suboptimal solutions.
\end{itemize} 

Both the multi-objective consideration and continuous charging rate control under physical limitation constraints introduce significant challenges on the convergence of RL.
To overcome these shortcomings, this paper formulates the joint pricing and continuous-multiple charging rates scheduling problem. The goal is to maximize both the charging station profit and reputation, considering the multi-objective nature and continuous charging rate control under physical limitations. The charging station's reputation is modeled with a pricing fluctuating penalty \cite{10.1086/508435,article},  and a weighted-sum based multi-objective optimization function is formulated. To handle total charging rate and deadline constraints, a linear programming-based safe layer is proposed to ensure feasibility with the continuous-multiple rates output of the deep neural network. However, the complexity of the continuous-multiple charging rates' action space, the trade-off among multiple objectives, and the sparse gradient induced by the safe layer make the training process extremely challenging. To address these convergence issues, we further propose a loss function to adaptively tune the entropy temperature.    In summary, the contributions of this paper can be outlined as follows. We 
\begin{itemize}
	\item Formulate the charging station pricing and continuous port-wise scheduling problem as a Markov Decision Problem (MDP) to jointly maximize the profit and reputation of the charging station.
	\item Propose a soft actor-critic-based reinforcement learning algorithm with a linear programming-based safe layer to handle total charging capacity and deadline finish constraints for the port-wise scheduling problem. We also propose a loss function to adaptively tune the entropy temperature to cope with the gradient vanish problem induced by safe layer. 
	%\item To improve the convergence for the complex joint optimization problem with large action space, we integrate  the proposed safe reinforcement learning algorithm with the entropy temperature tunning.
	\item Conduct  experiments  with real data, demonstrating that  the  proposed method can provide extra  25.45\%-52.20\% average JPR than the representative baselines.
\end{itemize}
\section{System Architecture}
We consider the operation of an EV charging station over a time horizon that is divided into $T$ time slots, equipped with $N$ charging ports.
EVs arrive at the charging station at random times. We denote
by $\mathcal{I}_t$ as the set of  EVs that arrive at the charging station at the
beginning of time slot $t$. 
Following~\cite{9653670}, we assume that the EV will not enter the charging station, if the EVs find the ports of the charging station are fully occupied.   
Let $t_i$ and $p_i$ denote the arrival time and the parking time of EV $i \in \mathcal{I}_t$, respectively. In particular, the charging station also determines an unique public charging price $r_t$ \$/kWh for all EVs that arrive at time $t$.   The EVs are assumed to be price sensitive. In response to $r_t$,  each EV $i$ sets its charging demand as $d_i = D_i(r_t)$ kWh, where $D_i(\cdot)$ is the demand response function of EV $i$.

Let $\mathcal{J}_t$ denote the set of EVs that parking at the charging station and have not finished their charging at the beginning of time $t$. For notation simplicity, we denote $\mathcal{K}_t=\mathcal{J}_t \cup \mathcal{I}_t$ to be the set of EVs yet
to be charged in time slot $t$.

At the beginning of time $t$, the charging station determines the charging
rate  of each EV $i \in \mathcal{K}_t$, denoted as $x_{i,t}$ kWh. The charging rates are constrained by
\begin{subequations}
	\begin{align}
	x_{i,t} &\le x^{\max}, \forall i, t, \\
	\sum_{i \in \mathcal{K}_t} x_{i,t} &\le U, \forall t, \\
		\sum_{t=t_i}^{t_i+p_i} x_{i,t} &\ge d_i, \forall i.
\end{align}
\end{subequations}
where $x^{\max}$ and $U$ denote the maximum in  individual and aggregator charging rates, respectively. Moreover, inequality~(1c) guarantees that
the charging demand of each EV is satisfied before its departure time.

As a result, at each time t the charging station collects a
payment of $\sum_{i \in \mathcal{I}_t} r_tD_i(r_t)$ from the EVs, and pays an electricity bill of $\sum_{i \in \mathcal{K}_t} c_tx_{i,t}$ to the utility company. The industrial electricity price charged to the charging station, i.e., $c_t$ \$/kWh, varies over time under the real-time pricing scheme. Due
to the uncertainty of the EV arrival process and electricity price, the charging station only knows the charging profiles of the EVs
that have already arrived. Likewise, only the past and current electricity prices are known. Overall, the profit of the charging station is computed as:
\begin{equation}
\sum_{t=1}^{T} \big[\sum_{i \in \mathcal{I}_t} r_tD_i(r_t)- \sum_{i \in \mathcal{K}_t} c_tx_{i,t}\big].
\end{equation}
% \balance

In addition to the effect of profit maximization, the reputation of charging station can be further affected for future EV arrivals \cite{9095998,9124670}. In this paper, we consider the %charging fluctuation and 
pricing fluctuation as the charging station reputation. %In particular, charging fluctuation may introduce additional battery degradation and thus is regard as a reputation penalty. The charging fluctuation penalty at time $t$ can be computed as:
% \begin{equation}
%  \sum_{i \in \mathcal{J}_t}|x_{i,t}-x_{i,t-1}|.
% \end{equation}
In particular, the different price offered in consecutive time slot may lead to
perceptions of fairness and to negative consequences for the
consumer, such as dissatisfaction, distrust , and lower intentions to
repurchase. The pricing fluctuation penalty at time $t$ can be computed as:
 \begin{equation}
	\lambda_1[r_{t}-r_{t-1}]^++\lambda_2[r_{t-1}-r_{t}]^+.
\end{equation}

Overall, the joint profit and reputation maximization problem can be formulated as:
\begin{subequations}
	\begin{align}
		\max_{\bm{r},\bm{x}} \quad &  \sum_{t=1}^{T} \Big[\sum_{i \in \mathcal{I}_t} r_tD_i(r_t)- \sum_{i \in \mathcal{K}_t} c_tx_{i,t}-	\lambda_1[r_{t}-r_{t-1}]^+\\
		&-\lambda_2[r_{t-1}-r_{t}]^+\Big]\\
		\textrm{s.t.} \quad &   0\le x_{i,t} \le x^{\max}, \forall i, t, \\
		&\sum_{i \in \mathcal{K}_t} x_{i,t} \le U, \forall t, \\
	&	\sum_{t=t_i}^{t_i+p_i} x_{i,t} \ge d_i, \forall i, \\
	& r_t \ge 0, \forall t.
	\end{align}
\end{subequations}

Given the future electricity price and EV arrivals, Problem (4) is a non-convex programming problem with linear constraints. However, in real world, it is difficult to predict the future realization of random EV arrival and electricity price. To capture the dynamics, we formulate the problem as a real-time sequential decision-making problem, i.e., a Markov Decision Process (MDP) problem. 
\section{Methodology}
At the beginning of each time slot $t$, the charging station determines the charging service price $r_t$ and charging scheduling $x_{i,t}$ for all EV $i$ parking at the charging station. The decision is based on observations from the newly EV arrivals, the residual EVs in the charging station, and the electricity price provided the utility company. In the following, we define the state $\mathcal{S}$, action $\mathcal{A}$, and reward function $\mathcal{R}$, respectively.
\subsubsection{State}
The charging station collects informative features as a state, providing evidence for the controller to take action.  In this section, we explicitly define informative features
in terms of fundamental properties of the problem.

First of all, the residual EVs from the last time slot is dependent with the current pricing and scheduling. Therefore, we include $\bm{Z} \in \mathbb{R}^{N\times2}$ as a feature, where $Z_{i,1}$ and $Z_{i,2}$ denote the residual demand and the residual parking time of the EV parking at charging port $i$. Notice that if there no EV parking at port $i$, we have $Z_{i,1}=0$ and $Z_{i,2}=0$. Moreover, the electricity price and the EV arrival share significant time correlation. Accordingly, we also include the past and current electricity prices $c_{t-h+1}$, $c_{t-h+2}$, $\dots$, and $c_t$, as well as the past and current EV arrivals $\mathcal{I}_{t-h+1}$,$\mathcal{I}_{t-h+2}$, $\dots$, and $\mathcal{I}_{t}$ as features. To help the controller compute the price fluctuation and how much parking port is idle, we further take the last service price $r_{t-1}$ and the number of parking EVs $|\mathcal{K}_t|$ into the state. Overall, the state at the beginning of time $t$ can be expressed as
\begin{equation}
\begin{aligned}
	s_t=
	\big[&\bm{Z},c_{t-h+1}, c_{t-h+2}, \dots, c_t,\\&\mathcal{I}_{t-h+1},\mathcal{I}_{t-h+2}, \dots, \mathcal{I}_{t},r_{t-1},|\mathcal{K}_t|\big].
	\end{aligned}
\end{equation}
\subsubsection{Action}
At the beginning of time $t$, the charging station determines the charging service price $r_t$ and the charging rate of each port $X_{i,t}$ in time $t$. The action is defined as 
\begin{equation}
	a_t=(r_t,\bm{x}).
\end{equation}
\subsubsection{Reward}
The reward function is designed according to the objective of the charging station management. In this paper, the reward at time $t$ is formulated as the weighted sum of the current profit, i.e., EV payment minus  bills to the utility, and the reputation penalty.
\begin{equation}
\begin{aligned}
	\text{reward}[i]=&
	\sum_{i \in \mathcal{I}_t} r_tD_i(r_t)- \sum_{i \in \mathcal{K}_t} c_tx_{i,t}\\&-	\lambda_1[r_{t}-r_{t-1}]^+-\lambda_2[r_{t-1}-r_{t}]^+.
	\end{aligned}
\end{equation}
\subsubsection{Soft Actor Critic Framework}
The primary objective is to determine the policy, denoted as $\pi$, that maximizes the long-term expected reward. Inspired by \cite{haarnoja2018soft}, to ensure continuous exploration by the agent, an entropy term,  $\mathcal{H}(\pi(a|s))=-\log\pi(a|s)$ , is incorporated into the reward.   Here, $\pi(a|s)$ denotes the probability of taking action $a$ given state $s$ under policy $\pi$.  The soft Q-value function is defined for the initial state-action pair  $(s,a)$ as follows: 
\begin{equation}
\begin{aligned}
  &Q^{\pi}(s,a)=\\
  &\mathbb{E}^{\pi}\{\sum_{t=0}^{\infty}\gamma^t\left[v_t(s_t,a_t)-\alpha\log\pi(a_t|s_t)\right]|s_0=s,a_0=a\}.
  \end{aligned}
\end{equation}
%In the actor-critic framework, the learning process alternates between policy improvement, i.e., actor module  {with parameter $\phi$, and policy evaluation, i.e., critic module with parameter $\theta$,  in the direction of maximizing 
Within the actor-critic framework, the learning process alternates between policy improvement (actor network with parameter $\phi$) and policy evaluation (critic network with parameter $\theta$), aiming to maximize 
$Q^{\pi_{\phi}}_{\theta}(s,a)$. Both the parameters are randomly initialized following the standard normal distribution at before time $0$.

In particular, the critic network takes $(s_t,a_t)$ as inputs and predicts the expectation of $Q^{\pi}(s_t,a_t)$, considering both the long-term reward and entropy. The soft Q-value is approximated as  $Q^{\pi}(s_t,a_t) \approx Q_{\theta}(s_t,a_t)$. %, where $\theta$ denotes the parameters of the critic network and is randomly initialized following the standard normal distribution at time $0$. 
Similar to the conventional value-based deep reinforcement learning algorithm, %the loss function of the critic network is the mean squared error (MSE) between $Q$ evaluated value $Q_{\theta}(s,a)$ and $Q$ target value $\hat{Q}_{\theta}(s,a)$, i.e.,
the soft Q-function parameters $\theta$ can be trained to minimize the soft Bellman residual
\begin{equation}\label{criticloss}\begin{aligned}
  &L_c(\theta)=\mathbb{E}_{s_t,a_t}\Big[\frac{1}{2}\big(Q_{\theta}(s_t,a_t)-\hat{Q}_{\theta}(s_t,a_t)\big)^2\Big],
  \end{aligned}\end{equation}
with
\begin{equation}
\begin{aligned}\label{qhat}
  &\hat{Q}_{\theta}(s_t,a_t)=v_t(s_t,a_t)+ \\&
\gamma \mathbb{E}_{a_{t+1}\sim \pi_{\phi}(s_{t+1})}\left[Q_{\bar{\theta}}(s_{t+1},a_{t+1})-\alpha\log\left(\pi_{\phi}\left(a_{t+1}|s^{h}\right)\right)\right].
  \end{aligned}
\end{equation} The update makes use of the target soft $\hat{Q}$ in (\ref{qhat}) with parameters $\bar{\theta}$ obtained as an exponentially moving average of the soft Q weights, contributing to training stability.
\subsubsection{Actor Network with Safe Layer}
  \begin{figure}[!t]
	\centering
	\includegraphics[width=0.99 \linewidth]{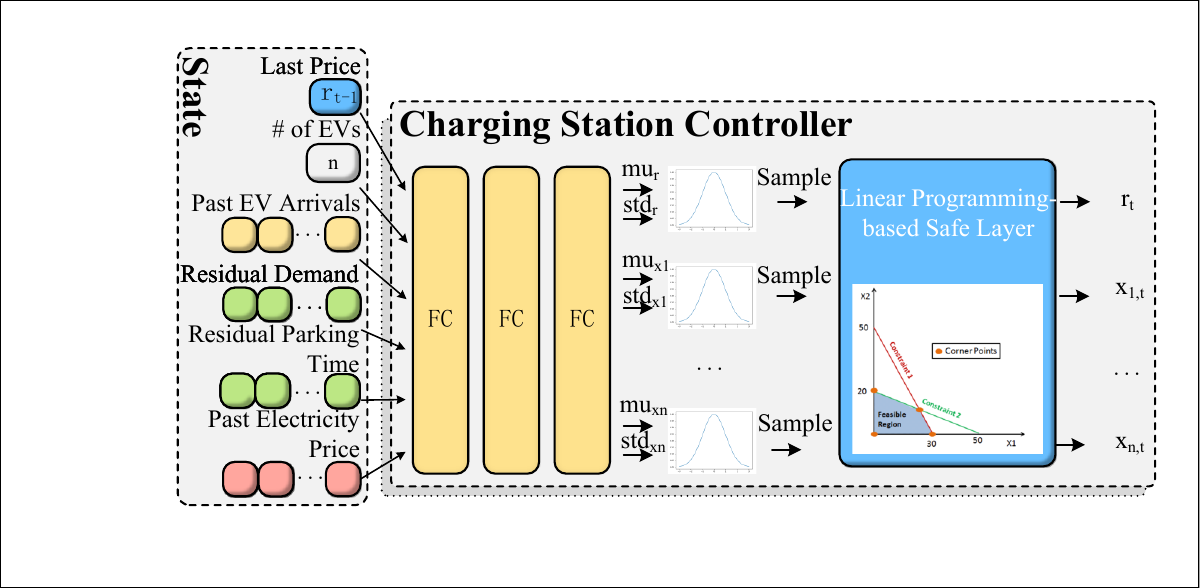}
	\caption{Actor Network with Safe Layer.}
	\label{sys}
	%\vspace{-0.4cm}
\end{figure}
As shown Fig.~\ref{sys}, the actor network takes $s_t$ as input, and return the charging service pricing and scheduling $(r_t,\bm{x})$. However, the conventional Gaussian action generator can not always ensure the feasibility of constraints (4d) and (4e). Accordingly, we propose a linear programming (LP) based safe layer to remap the output of Gaussian generator $\hat{\bm{x}}$ to satisfy the constraints. The LP is
\begin{subequations}
	\begin{align}
		\max_{\bm{x}} \quad & ||\bm{x}-\hat{\bm{x}}||_1\\
		\textrm{s.t.} \quad &   0\le x_{i,t} \le x^{\max}, \forall i, \\
		&\sum_{i \in \mathcal{K}_t} x_{i,t} \le U, \\
	&	x_{i,t}+\sum_{t=t_i+1}^{t_i+p_i} x^{\max} \ge d_i, \forall i.
	\end{align}
\end{subequations}
Overall, the optimal policy $\pi_\phi$ is learned by minimizing 
 \begin{equation}\label{actorloss}
   \begin{aligned}
   L_a(\phi)=\mathbb{E}_{s_t,a \sim \Big(\text{Standard}\big(\pi_{\phi}(s_t)+\epsilon\big)\Big)}\big[&\alpha \log\pi_{\phi}(a|s_t)\\&-Q^{\pi}_{\theta}(s_t,a)\big],
   \end{aligned}
 \end{equation}
where  $\text{Standard}(\cdot)$ is the standardization function to ensure that the summation of the probabilities is 1, and the noise vector $\epsilon\sim \mathcal{N}(0,1)$ is sampled from the standard normal distribution.
\subsubsection{Entropy Temperature Tunning}
To cope with the gradient vanish problem induced by safe layer mapping, we employ the following loss function to  adaptively tunning the entropy temperature $\alpha$:
\begin{equation}
	L_e(\alpha)=\mathbb{E}_s\big[-Q_{\theta}(s,\pi_{\phi}(s_t|\alpha))\big].
\end{equation}
\section{Simulations}
We base our simulations on the historic hourly data, including
the  electricity prices  of Shenzhen in China Southern Power Grid  and the number of vehicle
arrivals for Richards Ave station near downtown Davis. The DR
function is modeled as $D(r) = \beta_1 r + \beta_2$. The EVs are divided
into three types, namely emergent, normal, and residential uses.
The parameters of the three are listed in Table~\ref{tab:my-table}. The maximum charging rate of each charging port is set as 7 kWh, and the total charging capacity of the whole charging station is set as $5.6 \times N$, where $N$ is the total number of charging ports. The number of charging port varies from 3 to 7 in our simulations. The experimental setup is consistent with
 \cite{10323291}. Due to the page limit, please refer to \cite{10323291} for details.

This article targets a proof-of-concept study for joint profit and reputation optimization with continuous charging rate control under realistic physical constraints. Without loss of generality, the reputation parameters are set as $\lambda_1=1.0610$ and $\lambda_2=-0.2979$ \cite{huang2019comyco}, respectively.  we adopt the conventional implementation of the neural networks for both the actor and critic networks. In particular, we adopt 3 fully-connected layers  followed by a ReLU function for both actor and critic networks. The number of neurons are set as 256, 256, and $1+2N$, respectively. The length of history information is set as $5$. The learning rates are 0.0001 and 0.001 for actor network and critic network, respectively. The optimizer is adopted as Adam. We consider the following two representative baselines for comparison: 
\begin{itemize}
\item Fleet-Profit: Following the design in \cite{8888199}\cite{9444352}\cite{9764664}\cite{10323291}, this baseline adopts the same soft actor and critic networks with the proposed method, to determine the pricing and total charging rates of the station at each time slot. It utilizes  the least-laxity-first method~\cite{8888199}\cite{10323291} to allocate the total rate to each port.  The training reward is the profit in each time slot.
\item Fleet-JPR: In this baseline, we replace the reward in Fleet-Profit with the Joint Pricing and Reputation (JPR) defined in Eq. (4a-b).
\end{itemize}

\begin{figure}[!t]
\centering
	\begin{subfigure}[a]{0.48\textwidth}
	\centering
		\includegraphics[clip,  trim={0.1cm 0cm 1cm 0.5cm},width=0.9\linewidth]{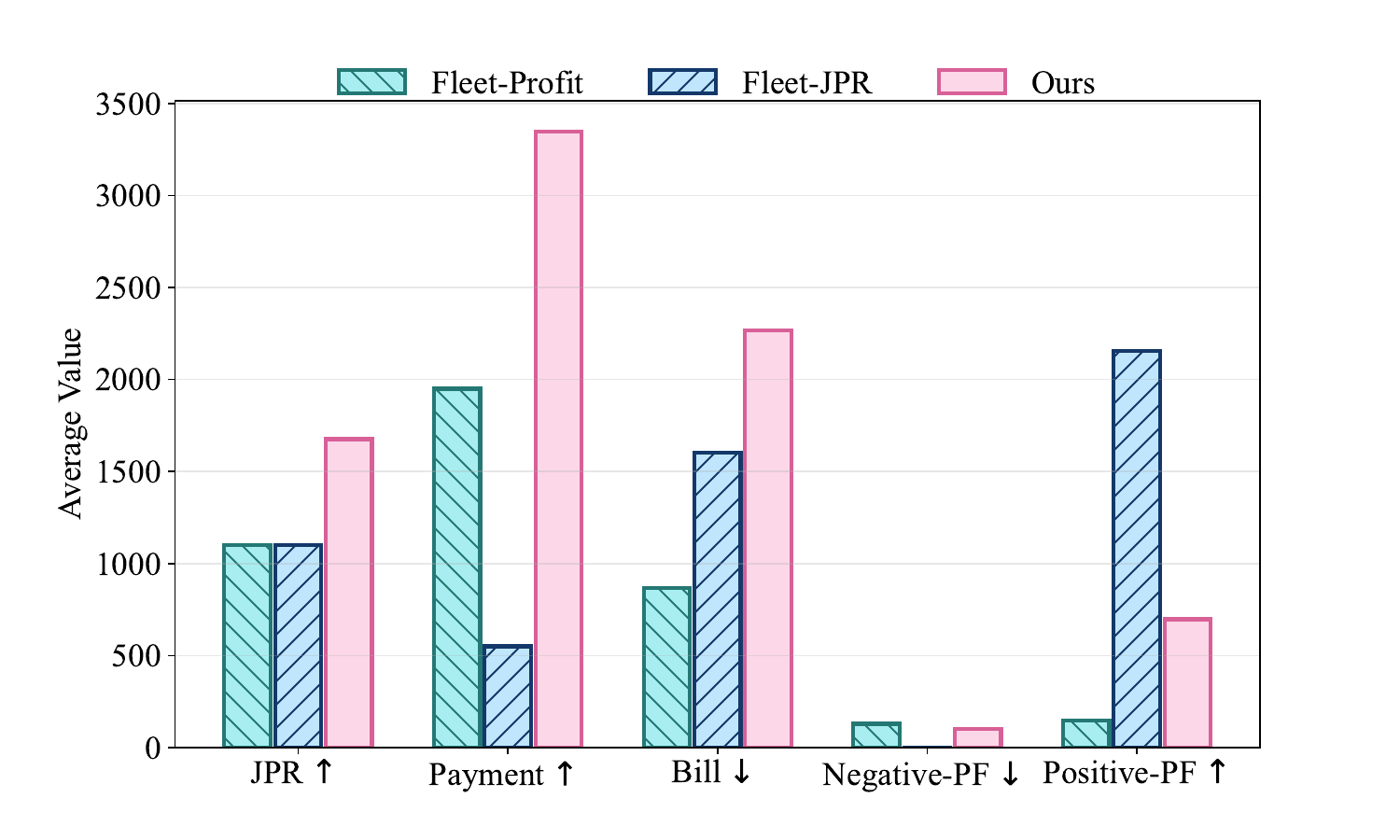}
		\caption{\small $\#$ of Ports: 5}
	\end{subfigure}
	\begin{subfigure}[b]{0.48\textwidth}
	\centering
		\includegraphics[clip,  trim={0.1cm 0cm 1cm 0.5cm},width=0.9\linewidth]{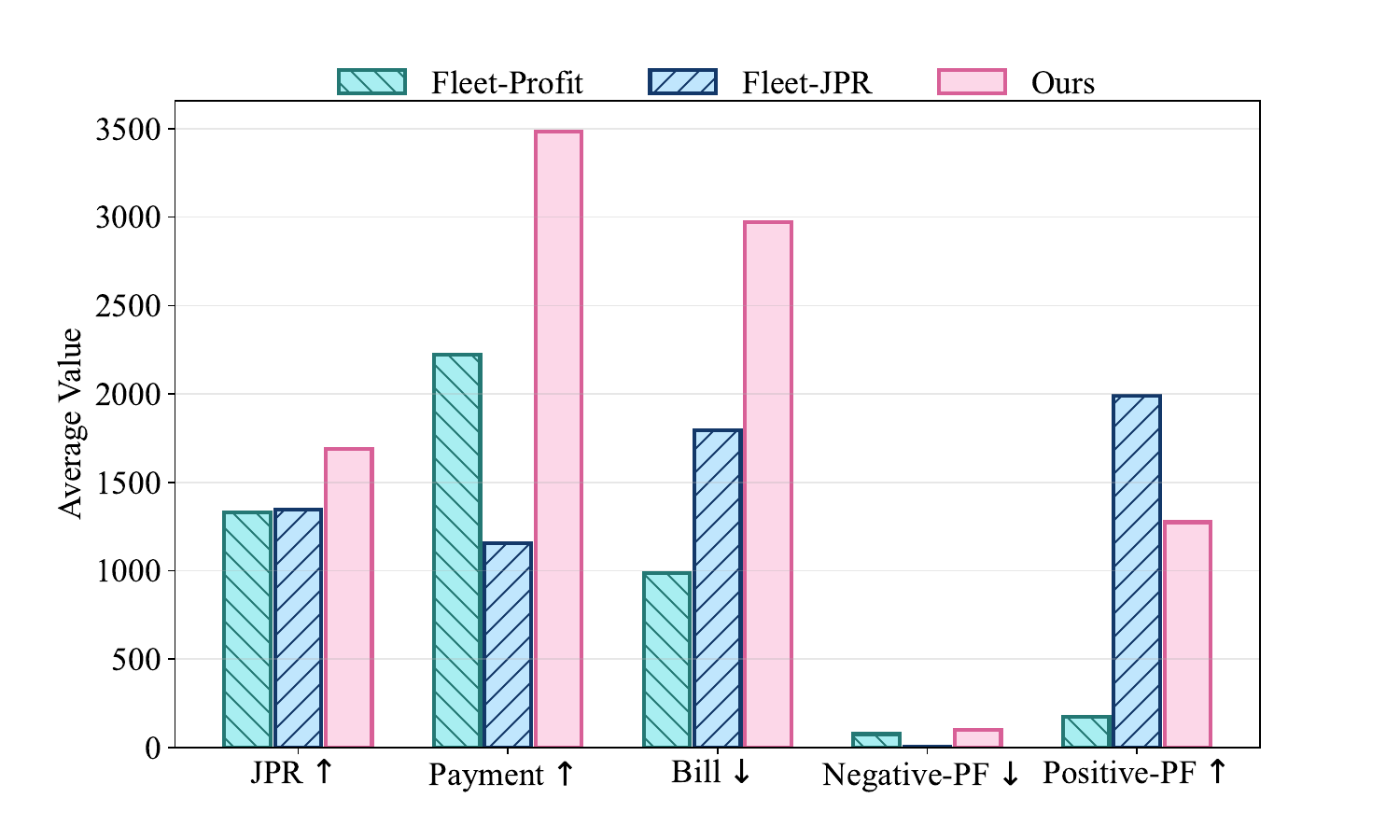}
		\caption{\small $\#$ of Ports: 6}
	\end{subfigure}
	\begin{subfigure}[c]{0.48\textwidth}
	\centering
		\includegraphics[clip,  trim={0.1cm 0cm 1cm 0.5cm},width=0.9\linewidth]{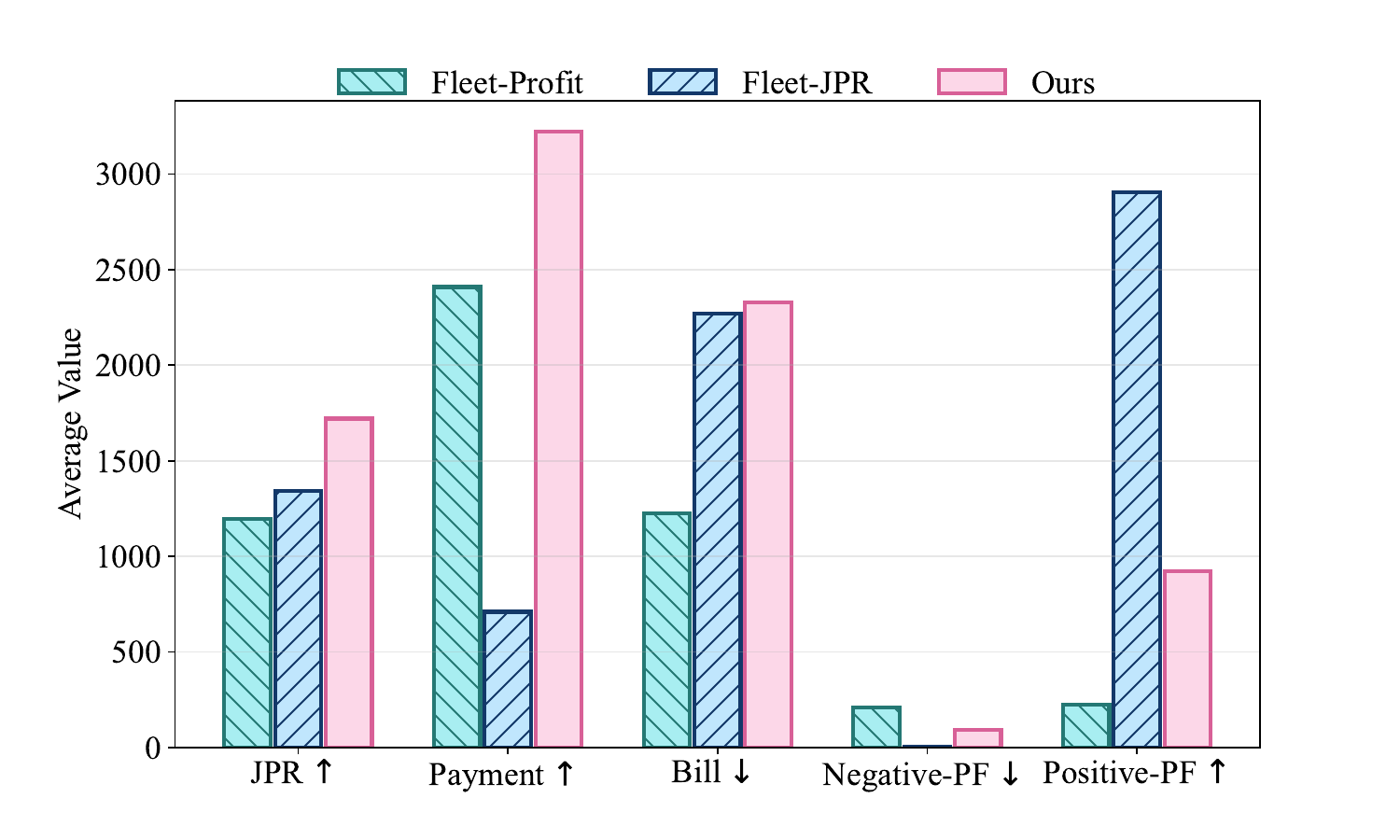}
		\caption{\small $\#$ of Ports: 7}
	\end{subfigure}
	\caption{\small Performance comparison versus different algorithms.}\label{fig:bar} 
\end{figure}

In the first experiment, we compare the JPR and its four components achieved by the proposed method and the baselines in Fig.~\ref{fig:bar}. The proposed method provides an additional  52.20\%, 25.45\%, and  28.12\% JPR when the number of charging ports is 5, 6, and 7, respectively, compared with Fleet-JPR. Notably, the proposed method accommodates more charging demand, maximizing both bills and charging service payments. In contrast, Fleet-JPR only marginally outperforms Fleet-Profit.  This validates the motivation of this work that the complexity of the JPR objective and  replacing JPR directly with the reward in existing work can lead to suboptimal solutions.

\begin{figure}[!t]
	\centering
	\includegraphics[width=0.85 \linewidth]{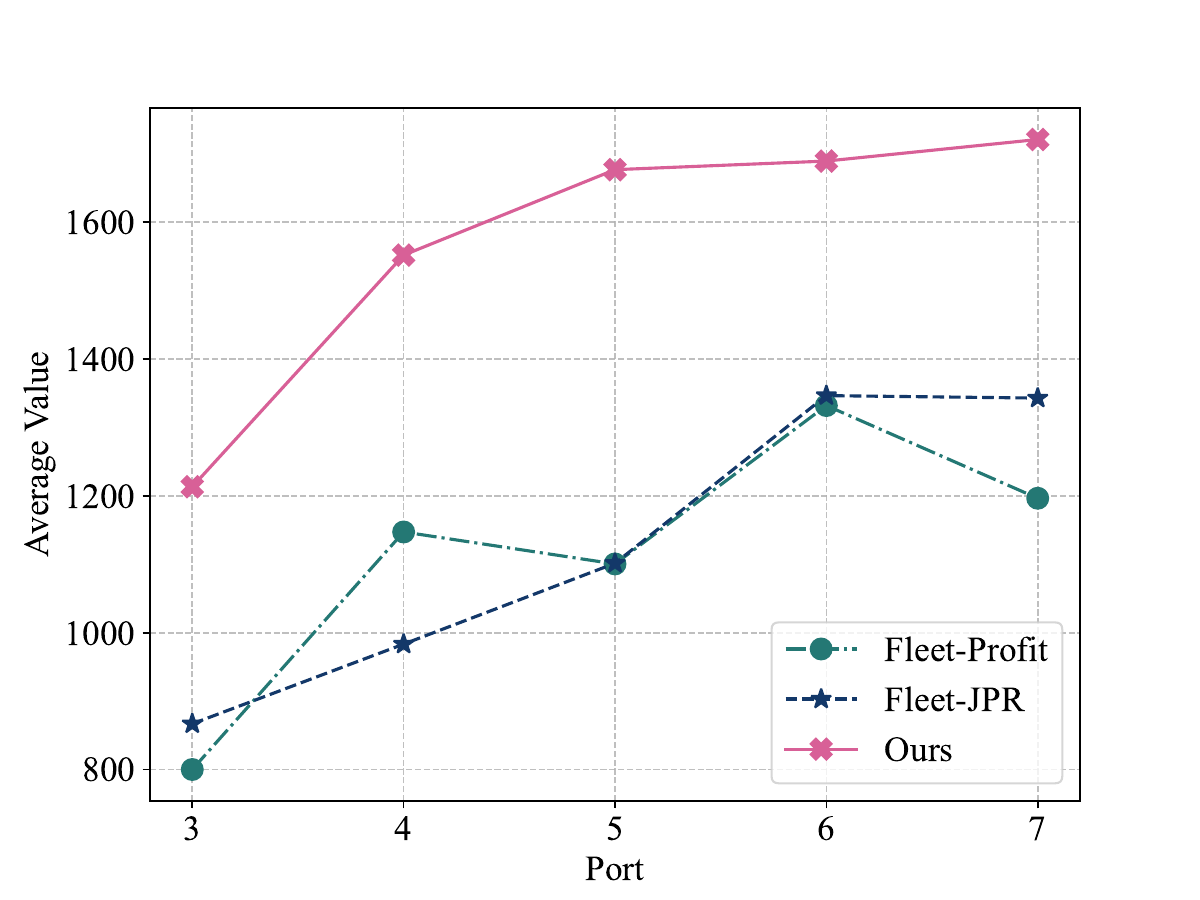}
	\caption{Performance comparison versus number of charging ports.}
	\label{port}
	%\vspace{-0.4cm}
\end{figure}

In the second experiment, we plot the JPR performances achieved by the proposed method and baselines when the number of charging ports increases from 3 to 7 in Fig.~\ref{port}. We can observe that the proposed method consistently outperforms the baselines under different number of ports. Compared with Fleet-JPR and Fleet Profit, the proposed method provide extra 40.70\% and 41.96\% average JPR. This performance gap is more significant when charging capacity is limited. This is because that when the charging capacity is limited, the charging station operation becomes more challenge. On the other hand, the Fleet-Profit baseline suffers performance fluctuation when the number of ports increases. This is because without considering the reputation, the reputation penalty achieved by Fleet-Profit becomes quite random. The Fleet-JPR achieves more stable performance by considering the reputation objective. However, the performance is limited, sometimes even worse than Fleet-Profit, due to the complex and conflict nature of the multi-objective target.
\begin{table}[!t]
\centering
\caption{Hyperparemeter setups.}
\label{tab:my-table}
\resizebox{0.8\columnwidth}{!}{%
\begin{tabular}{l|c|c|c}
\hline
User Type                            & Emergent & Normal & Residual \\ \hline
$\beta_1$ (Unit: 5kWh/CNY) & 2        & 10     & 24       \\ \hline
$\beta_2$ (Unit: 5kWh) & 4        & 12     & 32       \\ \hline
Deadline (Unit: 5mins)               & 3        & 6      & 12       \\ \hline
\end{tabular}%
}
\end{table}

\begin{figure}[!t]
	\centering
	\includegraphics[width=0.85 \linewidth]{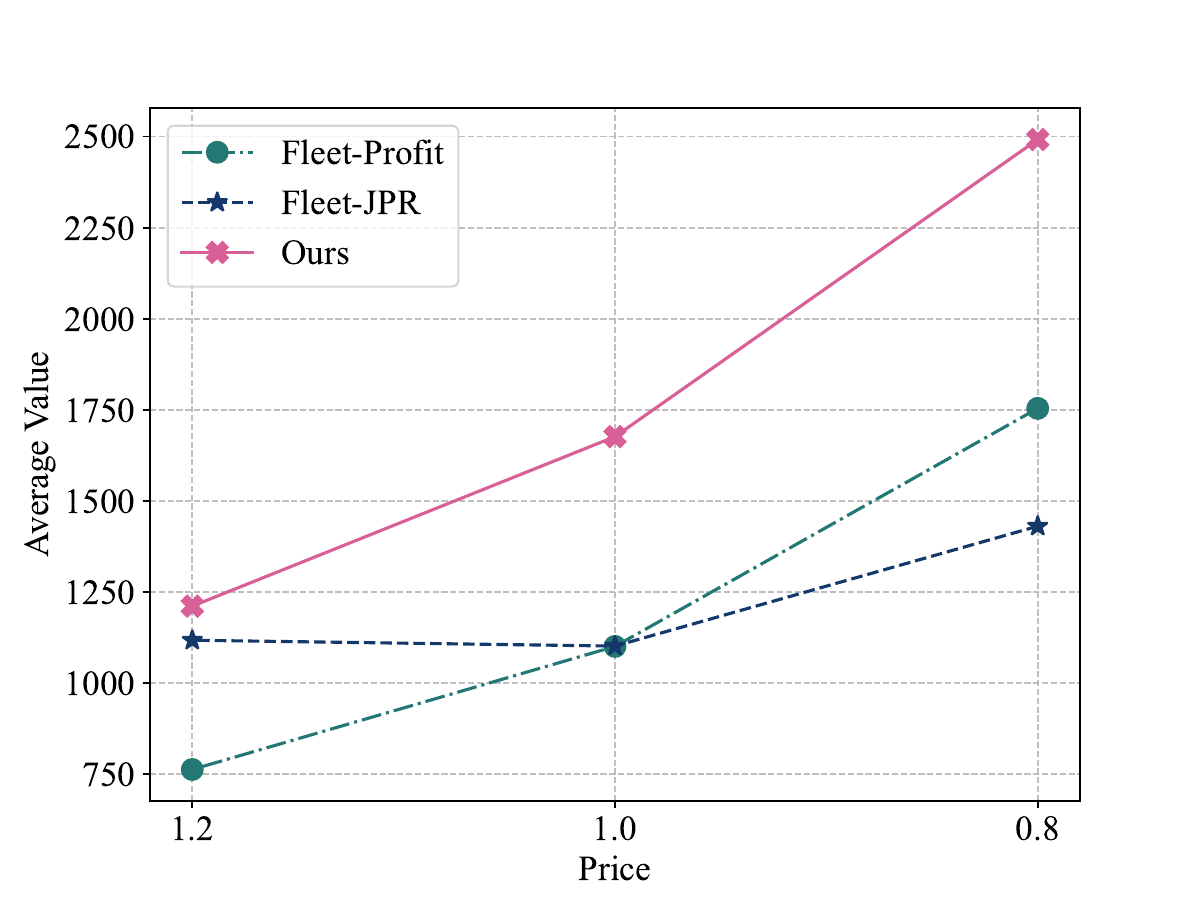}
	\caption{Performance comparison versus electricity prices.}
	\label{price}
	%\vspace{-0.4cm}
\end{figure}

In the third experiment, we further plot the JPR performance when the electricity price varies in Fig.~\ref{price}. In particular, we multiple a control factor (i.e., 0.8, 1.0, and 1.2) to the original electricity price from the South China Power Grid to simulate the performance under different electricity price. The proposed method consistently outperforms the baselines under different prices, with the performance gain widening as prices decrease. This highlights the proposed method's superiority in achieving robust scheduling and pricing schemes, even in complex scenarios, compared to baselines that diverge from optimal solutions when faced with large charging demand and complexity. 
\section{Conclusions}
In this paper, our study addresses the multi-objective challenges inherent in the charging station pricing and scheduling problem, aiming to optimize both the profit and reputation of the charging station concurrently. We cast the charging scheduling and pricing problem as a Markov Decision Process (MDP) and introduce a novel reinforcement learning (RL) algorithm to jointly maximize profit and reputation. To efficiently manage the constrained charging capacity in real-world scenarios, we tackle a port-wise continuous charging rate control problem with total charging capacity constraints. Overcoming the training challenges posed by this port-wise continuous charging and the complex multi-objective balancing problem, we integrate the soft actor-critic framework with a proposed linear programming-based safe layer and an entropy tuning technique.The experiments based on real-world data demonstrate that the  proposed method can provide extra  25.45\%-52.20\% average JPR than the representative baselines.
\bibliographystyle{IEEEtran}
\bibliography{IEEEfull} 
% \vfill

\end{document}